\documentclass[12pt,a4paper]{article}
\usepackage{graphicx}
\usepackage{geometry}
\usepackage{amsmath}
\usepackage{amssymb}
\usepackage{epstopdf}
\usepackage{hyperref}
\usepackage{chngpage}
\usepackage{subfigure}
\usepackage{float}
\usepackage{color,pictexwd,psfrag}
\restylefloat{table}
\newcommand{\figs}{draft_figs/}

\begin{document}

\begin{center}

\textbf{\LARGE{Modelling turbulent premixed flames using convolutional neural networks: application to sub-grid scale variance and filtered reaction rate}}

\vspace{0.25cm}

Zacharias M. Nikolaou \footnote[1]{Computation-based Science and Technology Research Centre (CaSToRC), The Cyprus Institute, Nicosia, 2121, Cyprus. z.nicolaou@cyi.ac.cy}, Charalambos Chrysostomou \footnotemark, Luc Vervisch \footnote[2]{CORIA - CNRS, Normandie Universit\'e, INSA de Rouen Normandie, France.}, Stewart Cant \footnote[3]{Department of Engineering, University of Cambridge.}

\end{center}

\begin{center}
\textbf{Abstract}
\end{center}

\noindent A purely data-driven approach using deep convolutional neural networks is discussed in the context of Large Eddy Simulation (LES) of turbulent premixed flames. The assessment of the method is conducted a priori using direct numerical simulation data. The network has been trained to perform deconvolution on the filtered density and the filtered density-progress variable product, and by doing so obtain estimates of the un-filtered progress variable field. Any non-linear function of the progress variable can then be approximated on the LES coarse mesh and explicitly filtered to advance the LES solution in time. This new strategy for tackling turbulent combustion modelling is demonstrated with success, for both the sub-grid scale progress variable variance and the flamelet filtered reaction rate, two fundamental ingredients of premixed turbulent combustion modelling.

%------------------------------------------------------------------
\newpage 

\section{Introduction}

Turbulent flows involve very disparate length and time scales \cite{batchelor_bookturb_1971,pope_book_2000}, and Direct Numerical Simulations (DNS) of the governing equations resolve all scales. This translates to a high computational cost which is prohibitive for most practical purposes. LES resolves instead only the largest, energy-containing motions of the flow \cite{smagorinsky_mwr_1963}, enabling flows of realistic size to be simulated. However, this loss of information manifests as a set of un-closed terms in the governing equations, and these terms require modelling in order to obtain a closed system. Traditional modelling efforts have focused on two fronts: (a) developing algebraic models, and (b) developing suitable transport equations. Such approaches have seen numerous applications throughout the years, and a multitude of different models have been developed \cite{gicquel_pec_2012}-\cite{sagaut_book_2001}. Traditional models have steadily been increasing in complexity in order to describe more demanding flows such as reacting, multi-phase etc. Furthermore, such models typically include many constants/parameters which require ``tuning" on a case-specific basis for an accurate simulation. These issues limit the application range of LES, but also undermine its credibility as a modelling tool. Given that increasingly more complex flows require simulating, it is unclear the extent to which traditional modelling approaches will be effective for future needs. 

Deconvolution-based modelling is an attractive alternative. Such methods do not explicitly depend on the flow regime, and do not involve any tunable parameters. The modelling of unresolved terms in LES is effectively an inverse problem. If an approximation of the original field $\phi$ can be obtained, namely $\phi ^*$, from its filtered counterpart $\bar{\phi}$, then any filtered function of $\phi$ namely $\overline{f(\phi)}$ can be modelled using $\overline{f(\phi ^*)}$, and the same applies for more than one variables. Such methods were first introduced in fluid mechanics research in the 1970s, in the works of Leonard \cite{leonard_adgeo_1974} and Clark \cite{clark_jfm_1979}. In later works, deconvolution was used for modelling purposes in non-reacting flows \cite{geurts_pof_1997}-\cite{stolz_pof_2001}, and for complementing traditional modelling efforts \cite{bose_pof_2014,locci_ftac_2016}. Deconvolution methods have also been employed for modelling purposes in reacting flows \cite{mathew_proc_2002}-\cite{domingo_cnf_2017}. More recently, different versions of constrained/un-constrained iterative algorithms were used for solving the inverse problem, and modelling a number of different terms in reacting LES with overall good results \cite{mehl_ctm_2017}-\cite{nikolaou_ftc_2018_01}. 

Data-driven methods, namely Deep Neural Networks (DNNs), are an ideal candidate for deconvolution in fluid mechanics. Despite numerous breakthrough applications of machine learning in a wide range of areas such as gene-profiling \cite{Khan_nat_2001}, speech recognition \cite{mikolov_atr_2011}, text-translation \cite{sutskever_procadvneur_2014}, decision-making \cite{Mnih_nat_2015,silver_nat_2016} etc. such methods have seen limited use for closure modelling in fluid mechanics, with most studies focusing in a more restrictive, non-deconvolution modelling context \cite{milano_jcp_2002}-\cite{maulik_jfm_2017}. In contrast to iterative algorithms, in data-driven approaches such as neural networks, an explicit knowledge of the filter kernel (user-defined) is not required. In a classic neural network, training data are used for obtaining the optimum node weights which minimise the error between the deconvoluted and original fields. Such a classic, single-layer, neural network was recently employed to reconstruct the velocity field in non-reacting flows and model the Reynolds stress terms \cite{maulik_jfm_2017}. Here, we propose using a deep Convolutional Neural Network (CNN). CNNs are a class of DNNs most commonly applied to analysing visual representations. In contrast to other state of the art methodologies, CNNs require comparatively less data pre-processing. Whereas in a traditional network the training is conducted on the raw data, the training in a CNN is performed on a set of ``features" which are extracted from the raw ``image". In the case of LES, the image consists of the three-dimensional filtered field which is to be deconvoluted. To extract the image features, CNNs perform a series of convolution and sampling operations on the raw input data using a number of different filtering kernels \cite{schmidhuber_nn_2015,lecun_2015_nat}. In contrast to classic networks, both the node (if any) weights and all the associated filter weights are optimised against the target results \cite{krizhevsky_procneural_2012}. As a result, CNNs are naturally suited for deconvolution. 

In this study, a high-fidelity DNS database is used in order to validate the method. A deep convolutional network is trained to perform deconvolution, for the original fields of two key variables, namely the density and the density-progress variable product. A stringent simulated LES mesh validation approach is employed \cite{nikolaou_prf_2018,
nikolaou_ftc_2018_01}, where the network training data size reduces with increasing filter width.
These approximations are then used to model the variance, which is a key parameter in many combustion models \cite{nikolaou_aaoaj_2018,nikolaou_ftc_2018_02}.
At this point it is also important to note that a priori assessments do not guarantee functionality of the models in actual LES. However, in contrast to a posteriori assessments where the influences of modelling and numerical errors are difficult to distinguish, a priori assessments minimise such issues and the performance of any model can be clearly evaluated.

\section{Description of the DNS database}  

The direct simulations have been conducted using the SENGA2 code \cite{senga2}. SENGA2 solves the compressible reacting Navier-Stokes equations for the conservation of mass, momentum, energy, and species mass fractions, using a 10th order finite difference scheme for interior points, and a 4th order Runge-Kutta scheme for the time-stepping,

\begin{equation}
\frac{\partial\rho}{\partial t} +
\frac{\partial\rho u_{k}}{\partial x_{k}} = 0\:,
\label{CONTINUITY}
\end{equation}

%the Navier-Stokes momentum equation:
\begin{equation}
\frac{\partial\rho u_{i}}{\partial t} +
\frac{\partial\rho u_{k}u_{i}}{\partial x_{k}} =
-\frac{\partial p }{\partial x_{i}}+
\frac{\partial \tau_{ki} }{\partial x_{k}}\:,
\end{equation}

%the internal energy equation:
\begin{equation}
\frac{\partial\rho E}{\partial t} +
\frac{\partial\rho u_{k} E }{\partial x_{k}} =
-\frac{\partial p u_{k}}{\partial x_{k}} -
\frac{\partial q_{k}}{\partial x_{k}}  +
\frac{\partial \tau_{km}u_{m}}{\partial x_{k}} \:,
\end{equation}

%and the species mass fraction equations:

\begin{equation}
\frac{\partial\rho Y_{\alpha}}{\partial t} +
\frac{\partial\rho u_{k} Y_{\alpha}}{\partial x_{k}} =
\dot w_{\alpha} -
\frac{\partial\rho V_{\alpha,k} Y_{\alpha} }{\partial x_{k}}\:.
\label{MASSFRAC}
\end{equation}

\noindent where $\alpha$ is the species identifier and usual notations have been otherwise introduced. A freely-propagating premixed fuel-air flame of a multi-component fuel is simulated, in a canonical inflow-outflow configuration. A detailed chemical mechanism was used, with 49 reactions and 15 species \cite{nikolaou_cnf_2013}. A turbulent fuel-air mixture flows from one end of the computational domain, burns, and the hot products leave from the other end of the domain. Table \ref{tbl:turb_param} lists the turbulence parameters for the direct simulations. The computational domain for cases A and B spans a ($L_x, L_y, L_z$) 14x7x7 mm domain, and for case C a 21x7x7 mm domain. $u_{rms}$ is the rms value of the fluctuating component of the incoming velocity field, and $l_\text{T}$ is the integral length scale in the reactant side. The turbulence Reynolds number is $Re_{\text{T}} =u_{rms}l_{\text{T}}/ \nu _r$, the Damkohler number is $Da=(l_{\text{T}}/u_{rms})/(\delta / s_{\text{L}})$ and the Karlovitz number is $Ka=(\delta/ \eta _k)^2$, where $s_{\text{L}}$ is the laminar flame speed, and the (diffusive) thickness $\delta=\nu _r/s_{\text{L}}$. The laminar flame thickness is defined as $\delta _{\text{L}}$=$(T_p-T_r)/\max(dT/dx)$ where $T_r$, $T_p$ are the reactant and product temperatures respectively. Three different turbulence levels were simulated spanning turbulence levels 3.18-14.04 as shown in Table \ref{tbl:turb_param}. \cite{nikolaou_cnf_2014,nikolaou_cst_2015}. These conditions place the flame in the distributed or broken reaction zones regime according to the classic combustion diagram by Peters \cite{peters_book_2000}. 

\begin{table}[ht!]
\centering
\begin{tabular}{l c c c c c c}
\hline
 Case & $u_{rms}/s_\text{L}$ & $l_\text{T}/{\delta}$ & $Re_\text{T}$ & $Da$ & $Ka$ & Use \\ [0.75ex]
\hline
A &3.18  &16.54 &52.66  &5.19  &1.39  & Training-validation \\
B & 9.00  &16.66 &150.05 &1.85 &6.62  & Training-validation \\
C & 14.04 &16.43 &230.69 &1.17 &12.97 & Testing             \\
\hline
\end{tabular}
\caption{Turbulent flame parameters for the training (A,B) and testing (C) DNS studies.}
\label{tbl:turb_param}
\end{table}

\begin{figure}[h!]
%\hspace{-1.0cm}
\subfigure[]{
\includegraphics[scale=0.30, trim=8.0cm 0.0cm 1.5cm 0.0cm]{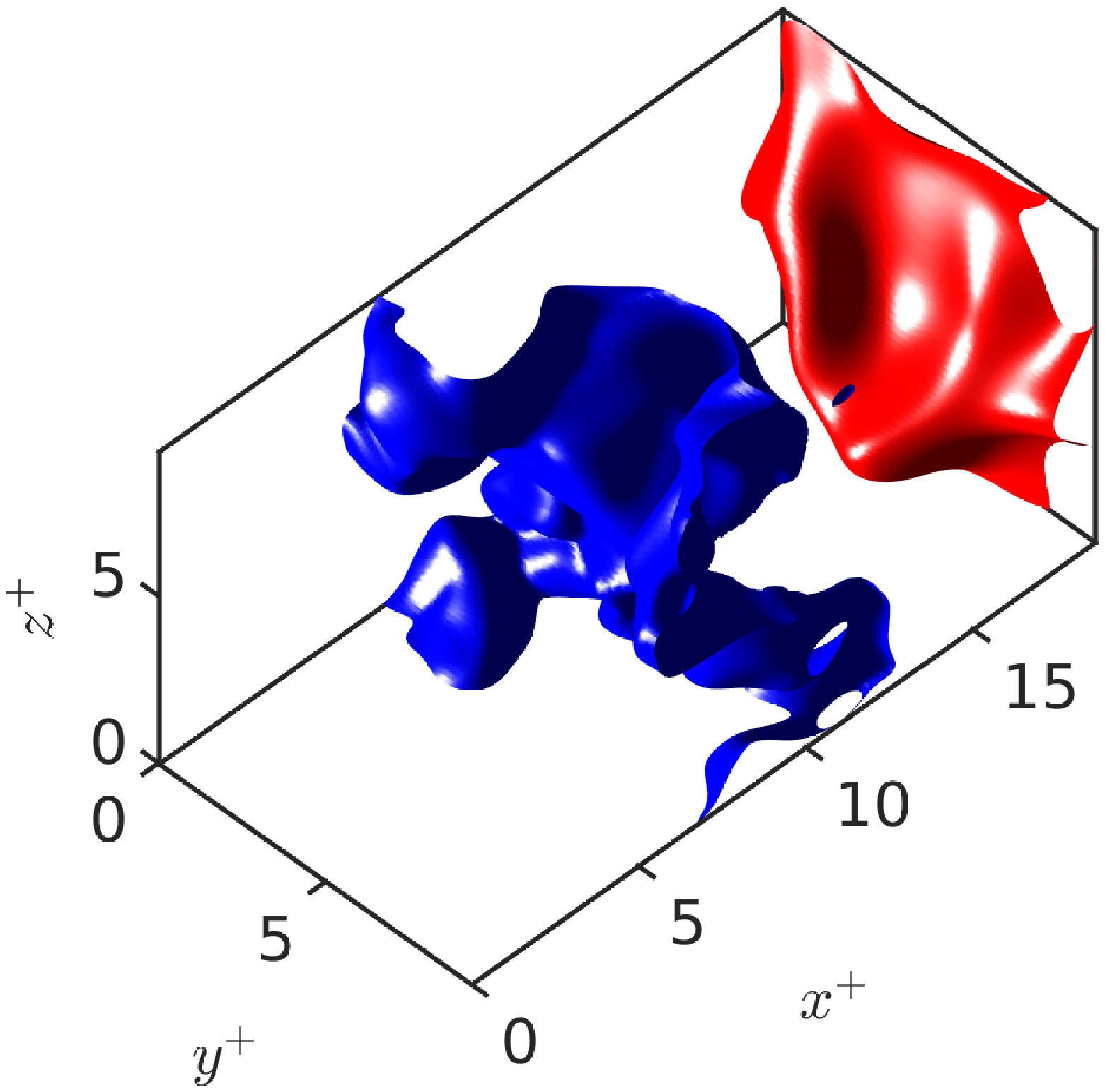}
}
%\hspace{1.0cm}
\subfigure[]{
\includegraphics[scale=0.30, trim=8.0cm 0.0cm 1.5cm 0.0cm]{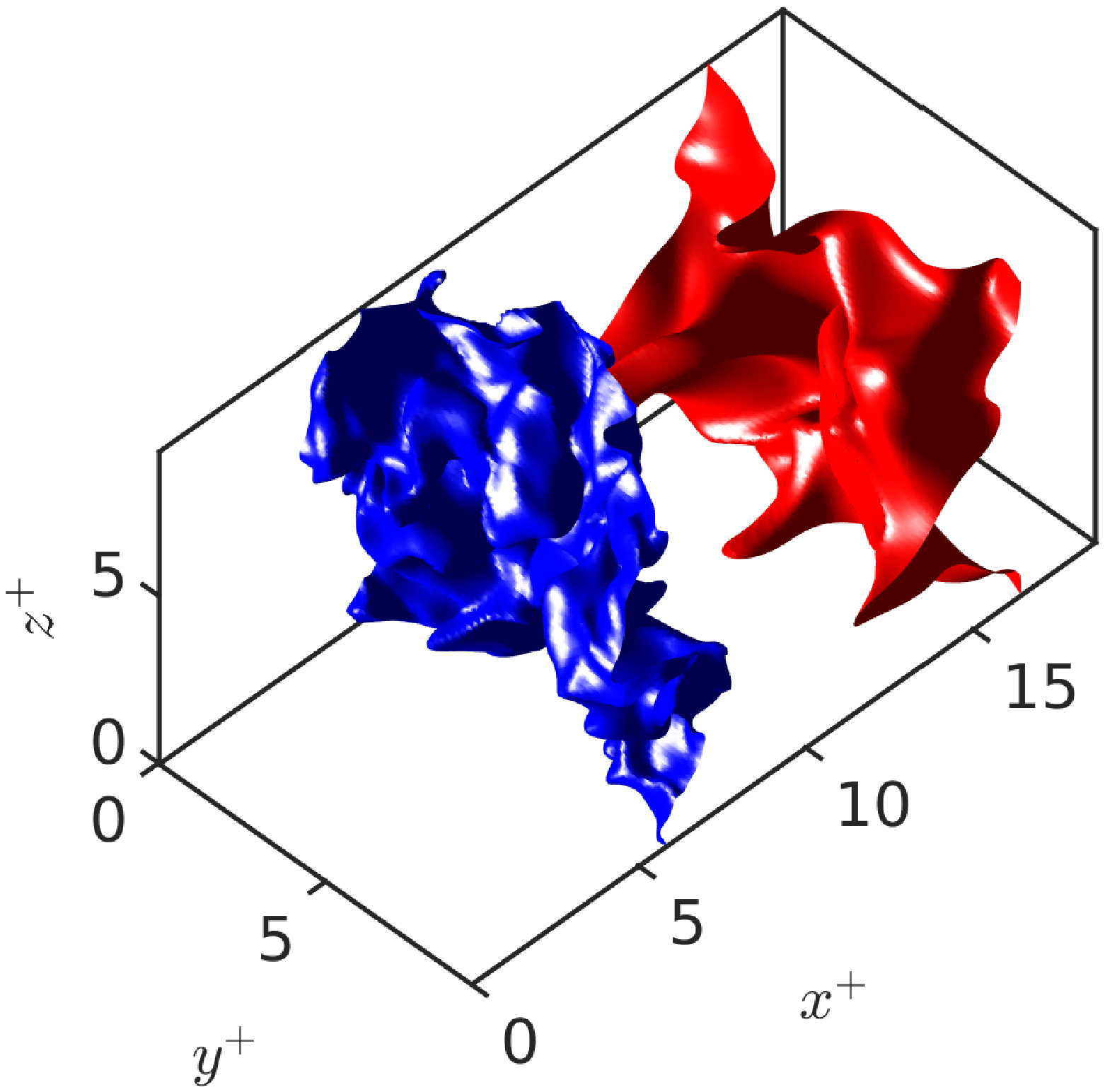}
}
\subfigure[]{
\includegraphics[scale=0.30, trim=8.0cm 0.0cm 1.5cm 0.0cm]{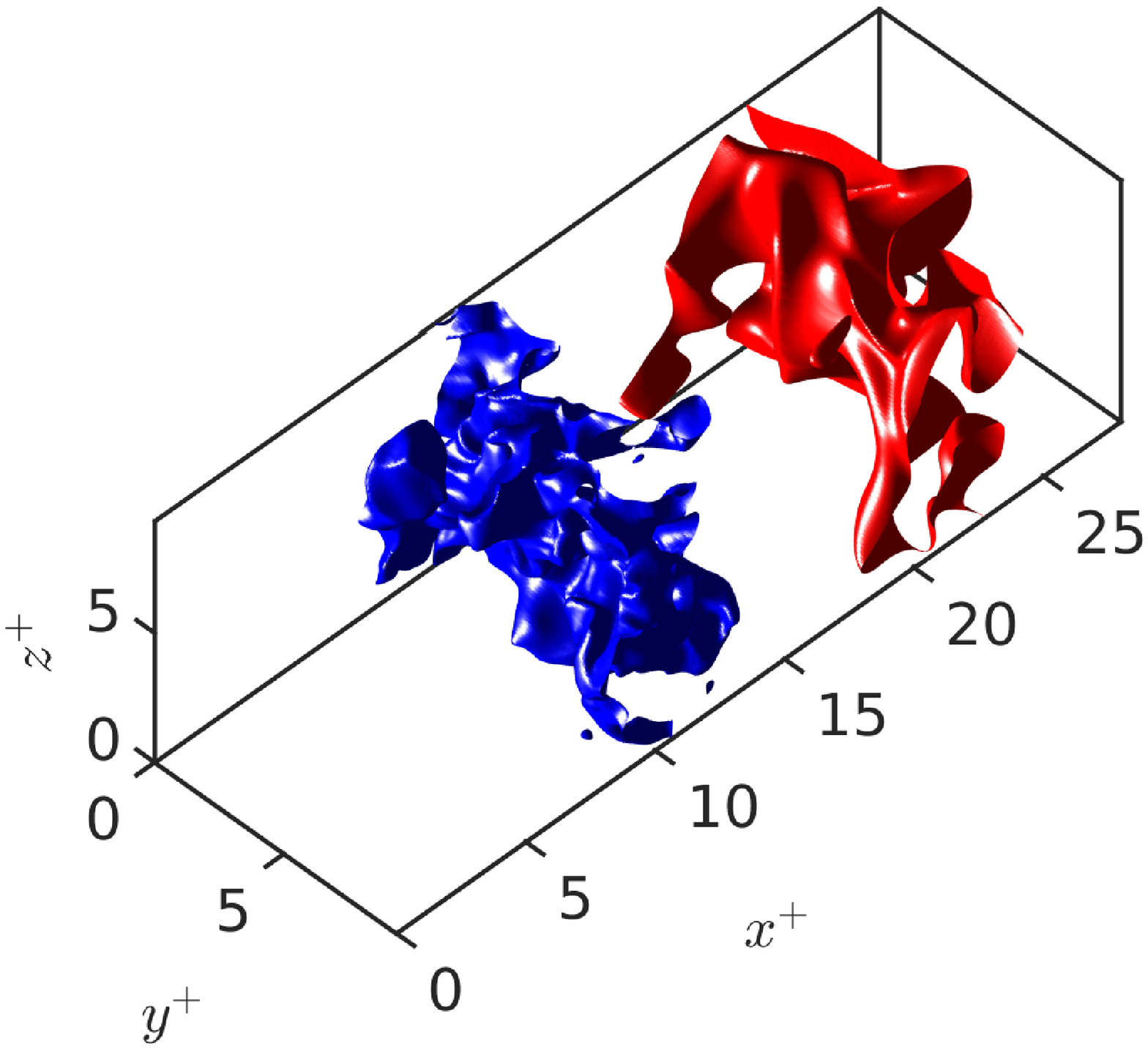}
}
\caption{Iso-surfaces of progress variable field ($c=0.1$-blue, $c=0.9$-red) on the DNS mesh for: (a) case A, (b) case B and (c) case C. Note the progressively finer-scale wrinkling for increasing turbulence level cases A to C.}
\label{fig:cdmesh}
\end{figure}

The low and intermediate turbulence level cases, A and B respectively, were used for training and validation of the network, while the highest turbulence level case C was used for testing the trained network. No data whatsoever from case C were used during the training or validation phase. Figure \ref{fig:cdmesh} (a)-(c) shows iso-surfaces of the progress variable field $c$ for the three different cases. Note that the axes are normalised using the laminar flame thickness $\delta _L$. In this study, the progress variable $c$ is based on temperature, $c(\underline{x},t)=(T(\underline{x},t)-T_r)/(T_p-T_r)$. The progress variable is an important parameter in reacting flow simulations, and is used to distinguish fresh ($c=0$) from burnt ($c=1$) gases. Figure \ref{fig:cdmesh} shows iso-surfaces corresponding to the leading ($c=0.1$, blue) and trailing ($c=0.9$, red) surfaces of the flame respectively. Turbulence decays in the $x$ direction which explains the heavier wrinkling of the $c=0.1$ iso-surface for each case. The mean turbulence level for case C is the highest, hence the wrinkling for both iso-surfaces is the heaviest amongst all cases. The training/testing scenario chosen thus presents a stringent test for the trained network, since the higher turbulence level case C exhibits a substantially more convolved flame surface geometry which is more difficult to deconvolute \cite{nikolaou_cnf_2014,nikolaou_cst_2015}. 
Further details of the simulations can be found in
\cite{nikolaou_cnf_2014,nikolaou_cst_2015}.

\section{Filtering}  

The DNS data have been explicitly filtered using a Gaussian filter. The filtered value of a variable $\bar{\phi}{(\underline{x},t)}$ is defined as,

\begin{equation}\label{eq:filtop}
\bar{\phi}(\underline{x},t)=\int\limits_{\underline{x}'=-\infty} ^{\infty}G(\underline{x}-\underline{x}') \phi(\underline{x}',t)d\underline{x}'
\end{equation}

\noindent The filter function is given by,

\begin{equation}
G(\underline{x})=\left( {\frac{6}{\pi {\Delta} ^{2}}} \right)^{\frac{3}{2}}  \exp\left ( -\frac{6\underline{x} \cdot \underline{x}}{{\Delta} ^{2}}\right )
\end{equation}

\noindent where $\Delta$ is the corresponding filter width. The laminar flame thickness is used as a basis for filtering at $\Delta ^+=\Delta / \delta_L$=1, 2 and 3. These choices correspond to filter widths which are significantly larger than the Kolmogorov length scale of the incoming turbulent field \cite{nikolaou_prf_2018,nikolaou_ftc_2018_01}. Favre-filtered variables are defined as,

\begin{equation}
\tilde{\phi}(\underline{x},t)=\frac{\overline{\rho \phi}}{\bar{\rho}}
\end{equation}

\noindent The DNS data have also been filtered for a period of more than one flame time $t_{fl}=t/(\delta_\text{L}/s_\text{L})$ when the flame was fully developed. Average quantities have also been time-averaged over the same period, in order to increase the statistical accuracy of the results. 

\begin{table}[h!]
\centering
\begin{tabular}{l c c c c}
\hline
$\Delta ^+$ & $N_x$ & $N_y$ & $N_z$ & $h$/ $l_T$  \\ [0.75ex]
\hline
DNS & 768 &384 &384 & 0.03 \\
1 & 74  &37  &37  & 0.35\\  
2 & 37  &18  &18  & 0.69\\
3 & 24  &12  &12  & 1.04\\
\hline
\end{tabular}
\caption{DNS and LES meshes for cases A  and B with $h / \Delta=0.25$.}
\label{tbl:mesh_A}
\end{table}

\begin{table}[h!]
\centering
\begin{tabular}{l c c c c}
\hline
$\Delta ^+$ & $N_x$ & $N_y$ & $N_z$ & $h$/ $l_T$ \\ [0.75ex]
\hline
DNS & 1632 &544 &544 & 0.02\\
1 & 112  &37  &37  & 0.21\\ 
2 & 56  &18  &18   & 0.42\\
3 & 37  &12  &12   & 0.64\\
\hline
\end{tabular}
\caption{DNS and LES meshes for case C with $h / \Delta=0.25$.}
\label{tbl:mesh_B}
\end{table}

In order to simulate an LES, the filtered data as obtained on the fine DNS mesh are sampled onto a much coarser LES mesh as per the approach described in \cite{nikolaou_ftc_2018_01}. Tables \ref{tbl:mesh_A} and \ref{tbl:mesh_B} list details of the DNS and LES meshes for each case. 
The choice of LES mesh size $h$ is based on the criterion derived in \cite{nikolaou_ftc_2018_01}. In particular, $h/ \Delta=$0.25 is used. It is important to note that the LES mesh is much coarser in comparison to the Kolmogorov length scale $\eta _k$: the ratio $h/\eta _k$ for cases A, B and C for the finest LES mesh ($\Delta ^+=$1) is 6.8, 14.9 and 20.7 respectively, and much larger for the coarsest LES mesh ($\Delta ^+=3$). As a result, small-scale information of the order of the Kolmogorov scale is not resolved on the simulated LES mesh. Details of the DNS and LES mesh size can be found in \cite{nikolaou_prf_2018,nikolaou_ftc_2018_01}. 

\section{Network training}

The training process begins by extracting the relevant variables of interest from the DNS database. These include the density $\rho$ and the product $\rho c$. A simulated LES mesh strategy is employed as per previous studies \cite{nikolaou_prf_2018,nikolaou_ftc_2018_01}. In contrast to traditional a priori studies which are conducted on the fine mesh, the simulated LES mesh approach presents a more stringent test since the coarsening of the mesh has a profound impact on the performance of any model \cite{nikolaou_prf_2018,nikolaou_ftc_2018_01}. A Gaussian spatial filter as indicated in the previous section, is used to filter a field of interest $\phi$. The solution variables from a typical LES would be the filtered density $\bar{\rho}$, and the filtered product $\overline{\rho c}$. These filtered fields would be obtained on a coarser LES mesh than typical meshes used in DNS. To simulate this, both $\bar{\rho}$ and $\overline{\rho c}$ as obtained on the fine DNS mesh are sampled onto the much coarser LES mesh using high-order Lagrange polynomials.

\begin{figure}[h!]
\centering
\hspace{2cm}
\includegraphics[scale=0.30, trim=8.0cm 0.0cm 1.5cm 0.0cm]{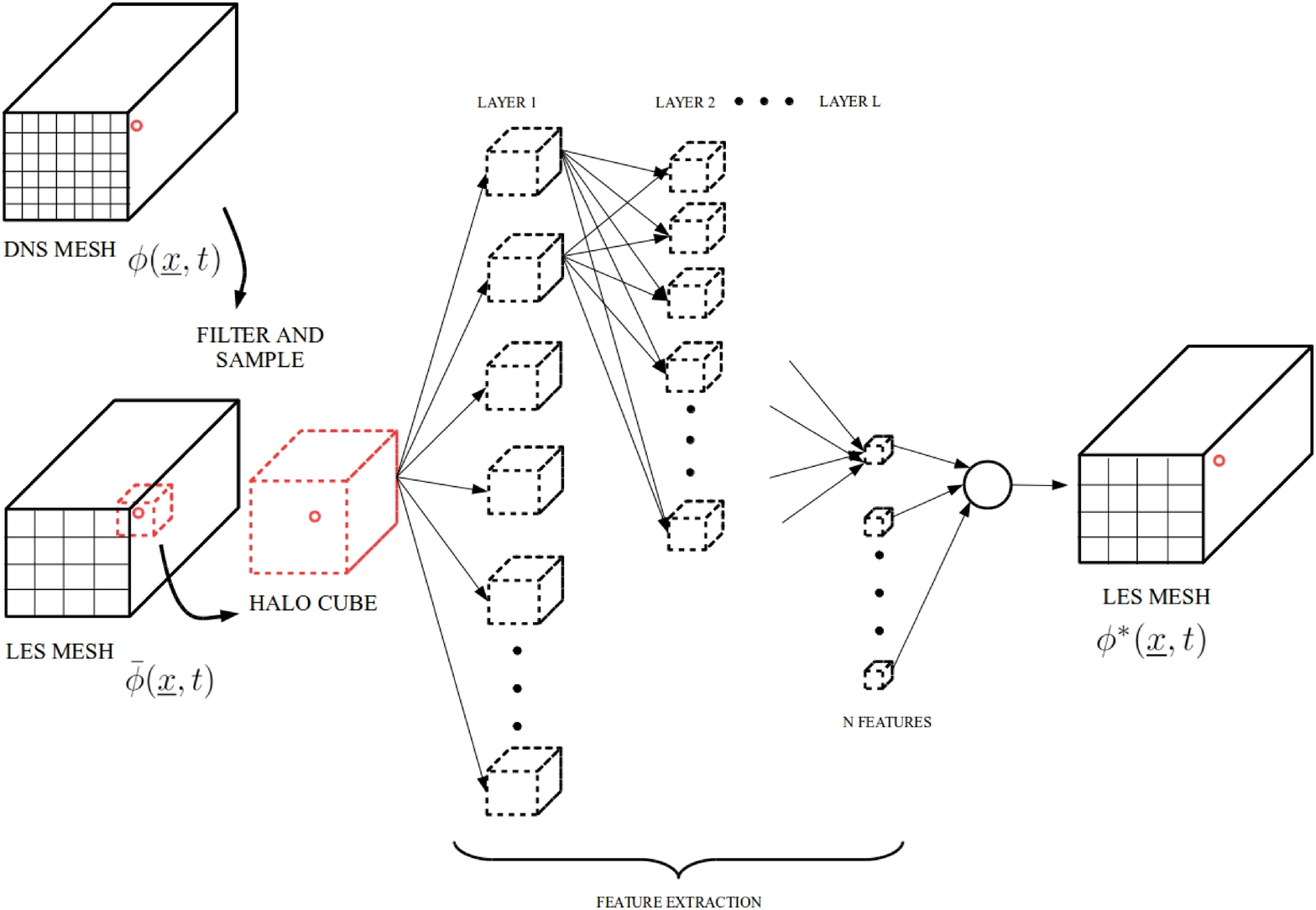}
\caption{The network training process: a field variable $\phi(\underline{x},t)$ on the fine DNS mesh is first filtered and then sampled onto the coarser LES mesh where only $\bar{\phi}(\underline{x},t)$ is known. The $\bar{\phi}$ in the halo cube around a point $\underline{x}$ are used to extract features to train the network. The size of extracted features (cubes) in each layer is reduced as we traverse the layers. Once trained, the network deconvolutes a given filtered field producing an approximation $\phi^*(\underline{x},t)$ of the original field on the coarse mesh.}
\label{fig:convnet}
\end{figure}

Figure \ref{fig:convnet} shows a diagram of the process. On the LES mesh, the only information available are the filtered values. The input to the network consists of the finite set of neighbouring filtered values in a ``halo" cube around a given point spanning the filter size. The halo cube in this study spans $11$ grid-points in each direction. Therefore, for each point $\underline{x}$ in space on the LES mesh and time $t$, the input layer to the network consists of an $11^3$ set of filtered values. The target output is the original field on the same mesh i.e. $\phi(\underline{x},t)$ as shown in Fig. \ref{fig:convnet}. Several convolution layers are employed. In each layer, a number of three-dimensional convolution kernels traverse the input data to produce a feature of the image. Each kernel may traverse all points in the input data e.g. all $11^3$ points for the input layer, or a finite set of sampled points-this is the preferred method in order to reduce the size of the extracted features and hence the training time. Each cube in Fig. \ref{fig:convnet} represents the set of features extracted for each convolutional kernel in the layer. Because of the sampling, the number of features reduces as we traverse the convolution layers-these are depicted as progressively smaller cubes in Fig. \ref{fig:convnet}. The weights of each kernel are initialised using random values. The aim of the sequence of convolution/sampling operations is to extract a final set of features which characterise the filtered field. The last layer consists of a one-dimensional array of features which are sent to a single node having a linear activation function. The optimisation problem in CNNs, is finding the associated convolution kernel weights in each layer which minimise the error between the deconvoluted field $\phi ^*$ and the target field $\phi$ on the LES mesh. A standard mean squared error between $\phi$ and $\phi ^*$ is used, which is usual practice. Details of the network structure are given in the Appendix. 

For each of the two variables, $\bar{\rho}$ and $\overline{\rho c}$, and for each filter width $\Delta$, a different network is trained and validated using data from cases A and B only. The structure of the network i.e. the number of layers and number of kernels is kept the same throughout, and only the weights of each layer are adjusted each time. The trained networks are then used to provide predictions of the deconvoluted fields namely $\rho ^*$ and $\lbrace \rho c \rbrace ^*$ for case C, on the LES mesh, for the same filter width as shown in Fig. \ref{fig:convnet}.

\section{Deconvoluted progress variable}

%These pictures are for sc01arch02
\begin{figure}[h!]
%\hspace{-1.0cm}
\subfigure[]{
\includegraphics[scale=0.35, trim=8.0cm 0.0cm 1.5cm 0.0cm]{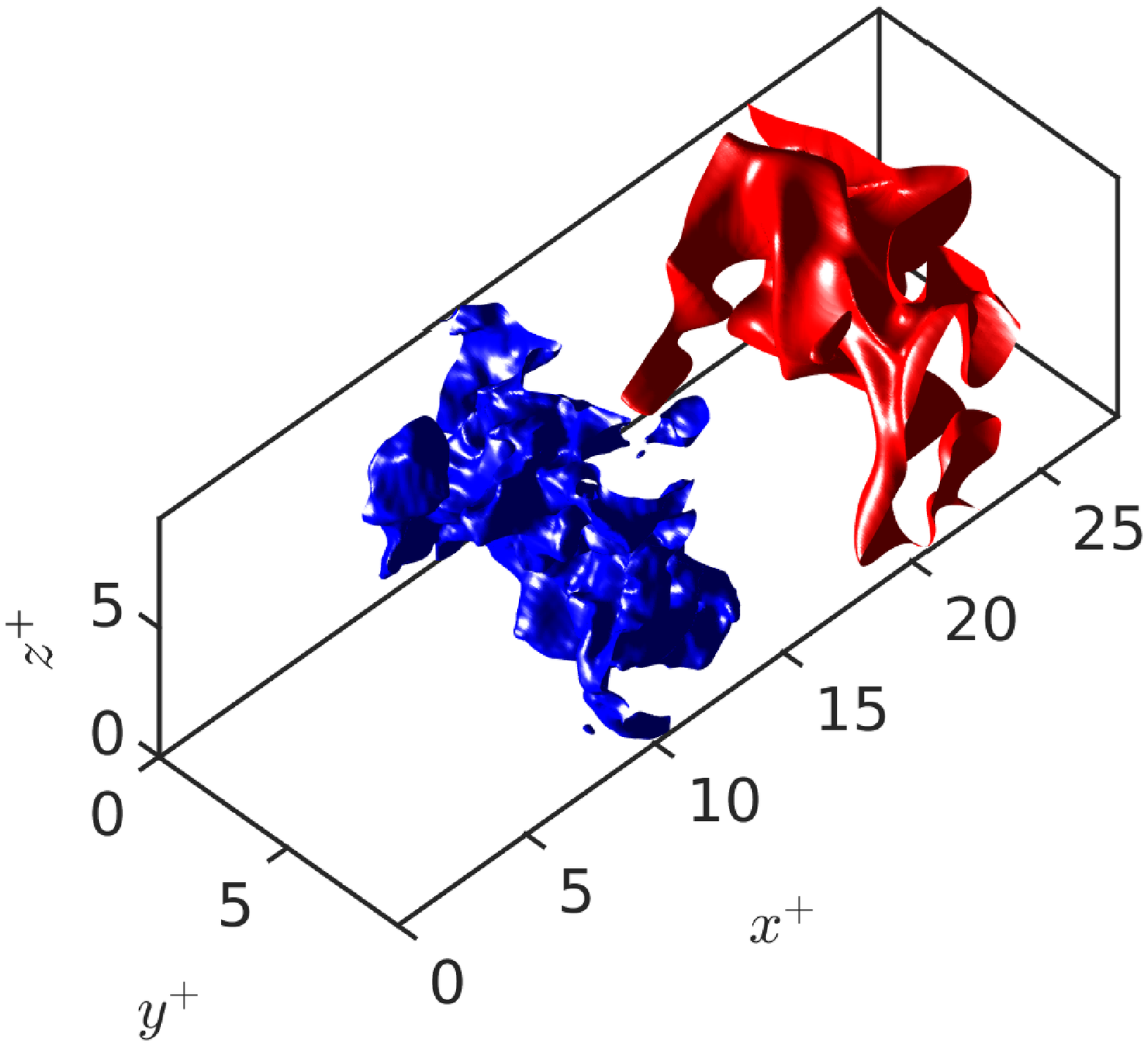}
}
%\hspace{1.0cm}
\subfigure[]{
\includegraphics[scale=0.35, trim=8.0cm 0.0cm 1.5cm 0.0cm]{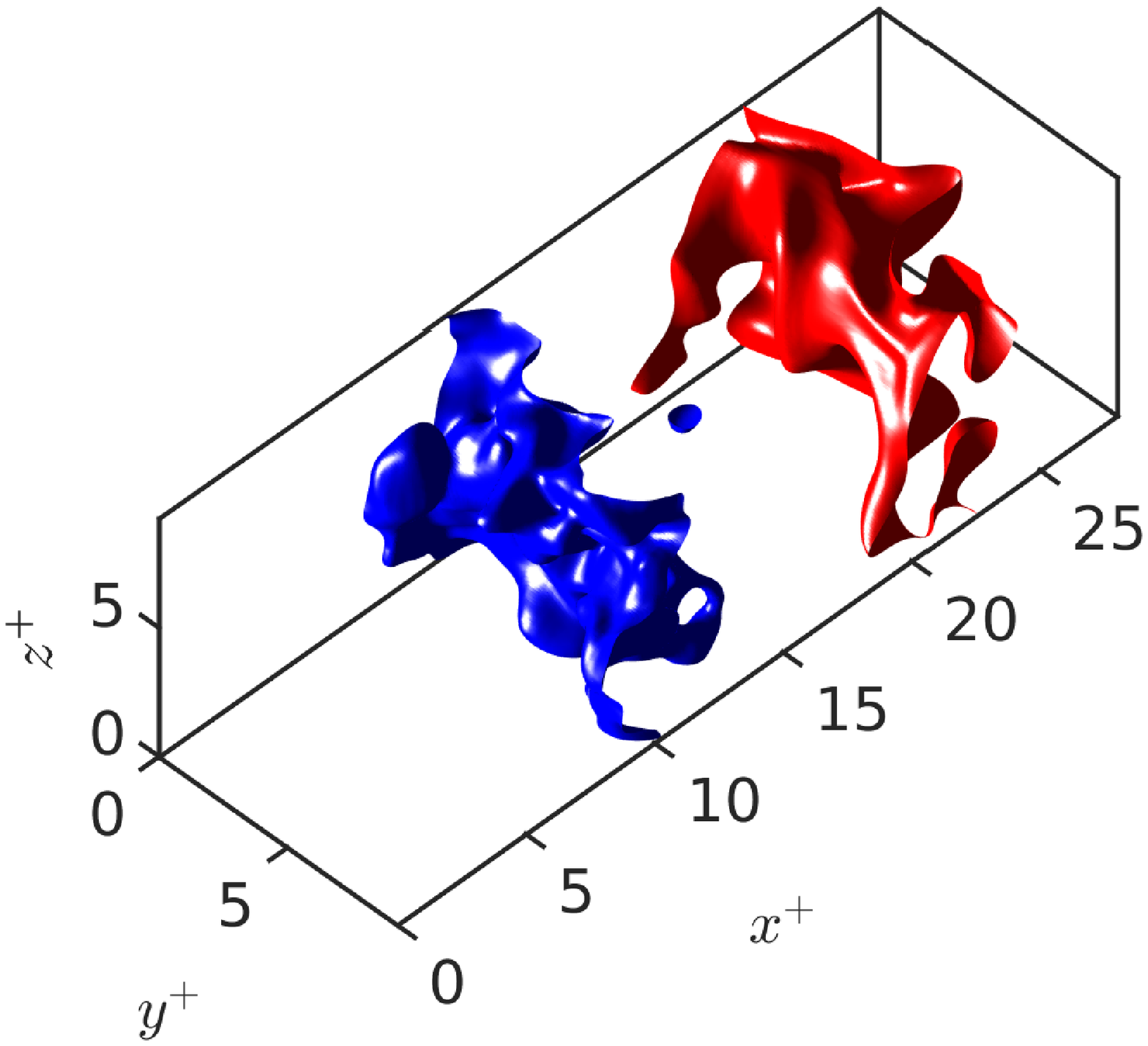}
}
%\hspace{1.0cm}
\subfigure[]{
\includegraphics[scale=0.35, trim=8.0cm 0.0cm 1.5cm 0.0cm]{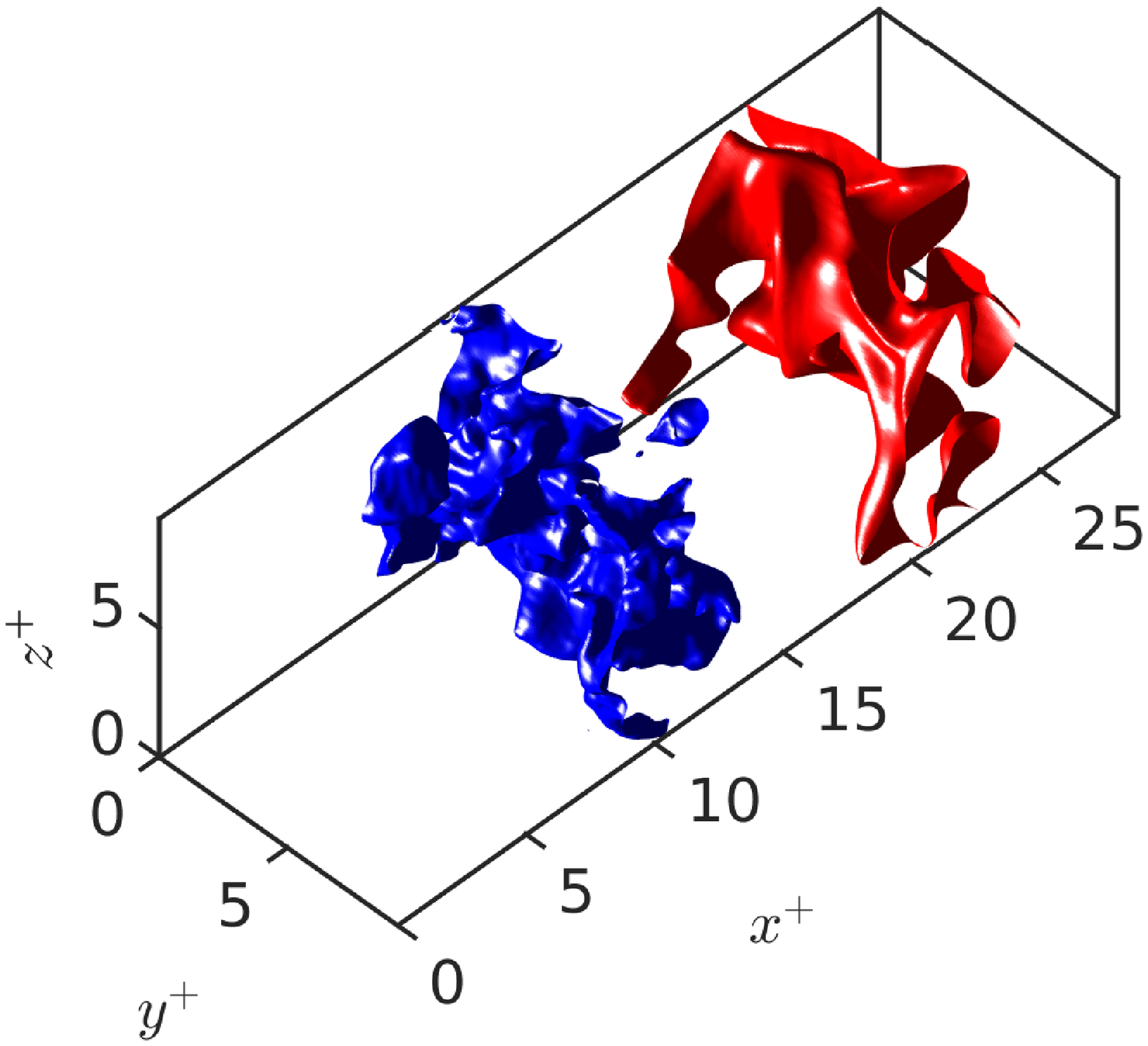}
}
\caption{Iso-surfaces of progress variable field (0.1-blue, 0.9-red) on the LES mesh for $\Delta ^+=$1: (a) original field $c$, (b) filtered field $\tilde{c}$, and (c) deconvoluted field $c^*$ using the trained CNN. Note the pronounced loss of small-scale information due to the the filtering in (b) on the reactant (blue) side where turbulence is more intense, and the recovery of the small-scales in the deconvolution step in (c).}%sc01arh02
\label{fig:deconv3D}
\end{figure}

Figure \ref{fig:deconv3D} (a)-(c) shows a comparison between the progress variable fields: the original field $c$, the Favre-filtered field $\tilde{c}=\overline{\rho c}/ \bar{\rho}$, and the deconvoluted field which is obtained using the deconvoluted fields of the density-progress variable product, and density i.e. $c^*=\lbrace \rho c \rbrace ^*/ \rho ^*$. The filtered field in Fig. \ref{fig:deconv3D} (b) corresponds to a filter size of $\Delta ^+$=1. Two different iso-surfaces are shown, corresponding to the leading and trailing surfaces of the flame. From a visual inspection alone, Figs. (a) and (c) are in good agreement. The network is able to recover much of the small-scale features both on the leading edge of the flame where the intense turbulence induces high wave-number flow components (small-scale wrinkling), but also on the product side where smaller wave-numbers are found as a result of the decaying turbulence. 

In order to better quantify the performance of the CNN-based deconvolution, mean percentage errors have also been calculated. These are defined using,

\begin{equation*}
e(x,y)=\frac{100}{N} \cdot \sum _{i,j,k,t}\frac{|x_{i,j,k,t}-y_{i,j,k,t}|}{|y_{i,j,k,t}|}
\end{equation*}

\noindent where $x$ and $y$ are the predicted and target variables respectively, and $N$ is the total number of sample points. In order to increase the statistical accuracy of the results, the errors are also time-averaged over a period where the simulation approaches a statistically stationary state. Table \ref{tbl:errors_sc01ar02} shows the errors for the deconvoluted variables $\rho ^*$ and $\lbrace \rho c \rbrace ^*$, and also for the derived quantity $c ^*$ for each value of the filter width. Note that in order to alleviate any biases from low and high values of $c^*$ which are proportionately much more since the flame actually occupies only a thin region in space, the error for $c^*$ is conditioned as indicated in Table \ref{tbl:errors_sc01ar02}. This conditioning ensures that points well within the flame thickness are considered. For the smallest filter width, the results are impressive: for all three variables the error is well below 5\%.  For increasing filter widths the errors increase as expected due to the reduction in the training data size. Overall however, the errors are reasonable and of the same order of magnitude as those found using iterative methods \cite{nikolaou_prf_2018}.

%Binning even!
\begin{table}[h!]
\centering
\begin{tabular}{l c c c}
\hline
Errors-CNNs & $\Delta ^+$=1 & $\Delta ^+$=2 & $\Delta ^+$=3  \\ [0.75ex]
\hline
$e(\rho ^*, \rho)$                      & 0.4374  & 1.3470  & 2.4747  \\
$e( \lbrace \rho c \rbrace ^*, \rho c)$ & 1.1389  & 5.7941  & 9.9235   \\
$e(c^*, c)| 0.2<c<0.8$                  & 1.5450  & 7.1040  & 12.4815   \\
\hline
\end{tabular}
\caption{Percentage errors between deconvoluted and original fields for testing case C.}
\label{tbl:errors_sc01ar02}
\end{table}

In a further test of the CNN-based deconvolution, the trained network as obtained from the filtered data for $\Delta ^+ =1$, is used in order to provide predictions for $\Delta ^+ = 2$, and 3. The aim of this test is to examine the sensitivity of the training approach to the filter width variation which is a key parameter in LES. Table
\ref{tbl:errors_sc01ar02_interfilter} shows the percentage errors for $\rho ^*$, $\lbrace \rho c \rbrace ^*$ and $c^*$ respectively for this testing scenario. The errors for both filter widths are found to be similar to the errors obtained by using the filter-specific-trained networks. In fact, the errors for $\rho ^*$ and $\lbrace \rho c \rbrace ^*$ are actually somewhat smaller due to the larger training data size for $\Delta ^+ =1$. These results indicate that the CNN-based deconvolution method is relatively insensitive to variations in filter width. This is in contrast to using classic, single-layer networks \cite{maulik_jfm_2017}, but also using classic modelling approaches, most of which have strong dependencies on the filter width. Essentially, these results imply that a single network trained on a relatively large data set for a small filter width is enough to deconvolute fields corresponding to larger filter widths. This result is directly related to the nature of convolutional networks.

%Binning even!
\begin{table}[h!]
\centering
\begin{tabular}{l c c}
\hline
Errors-CNNs & $\Delta ^+$=2 & $\Delta ^+$=3  \\ [0.75ex]
\hline
$e(\rho ^*, \rho)$                       & 1.5429  & 2.5537  \\
$e( \lbrace \rho c \rbrace ^*, \rho c)$  & 4.9562  & 8.9300   \\
$e(c^*, c)| 0.2<c<0.8$                   & 7.5100  & 12.6407    \\
\hline
\end{tabular}
\caption{Inter-filter testing percentage errors: using the weights for trained network on data from filter having $\Delta ^+$=1 to obtain estimates of $\rho$ and $\rho c$ for filters having $\Delta ^+$=2 and 3. The network is able to perform equally well, even though it was not trained on filtered data for filters having $\Delta ^+$=2 and 3.}
\label{tbl:errors_sc01ar02_interfilter}
\end{table}
%C-sc01-architxtr02 for interfilter testing.

\section{Modelling the variance}

\begin{figure}
\centering
\includegraphics[scale=0.35, trim=9.0cm 0.0cm 0.0cm 0.0cm]{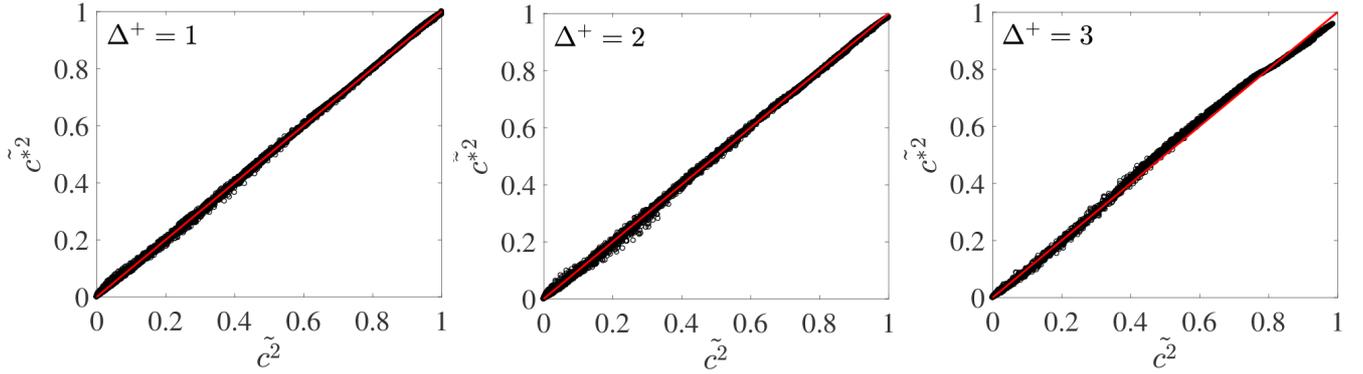}
\caption{Scatter plot of $\widetilde{{c^*}^2}$ as obtained using the convolutional network against the actual value-the red line corresponds to $y=x$.}%sc01arh02
\label{fig:c2}
\end{figure}

The aim of this section is to illustrate how CNN-based deconvolution can be directly applied for  modelling purposes in LES. The progress variable sub-grid variance, $\sigma ^2$, is an important quantity in the modelling of turbulent flows \cite{peters_book_2000}. In the context of LES, the variance is defined as,

\begin{equation}
\sigma ^2=\widetilde{c^2}-\tilde{c}\tilde{c}=\frac{\overline{\rho c^2}} {\bar{\rho}}-\tilde{c}\tilde{c}
\end{equation}

\noindent and is widely used in classic modelling approaches such as Conditional Moment Closure (CMC), in flamelet methods where a pdf is presumed, and in tabulation methods. The variance calculation requires the filtered product $\widetilde{c^2}$ which is not typically solved for in the LES and is an un-closed term. To this end, a number of different models were developed in the literature of increasing complexity often involving a number of different LES solution variables and their spatial gradients \cite{cook_pof_1994}-\cite{domingo_cnf_2008}. In this section, we illustrate how CNN-based deconvolution can be used in order to model the variance. In contrast to classic models, the deconvolution-based method is simple and straightforward to implement. The variance is simply calculated using the deconvoluted fields as, 

\begin{equation}
\sigma ^2=\frac{\overline{ \lbrace \rho c \rbrace ^* \lbrace \rho c \rbrace ^* /\rho ^*}}{\bar{\rho}}-\tilde{c} \tilde{c}
\end{equation} 

\noindent i.e. using the deconvoluted fields in order to obtain an estimate of the function $\rho c^2$. Once this estimate is obtained, explicit filtering is then used on the deconvoluted field (on the LES mesh) in order to obtain the filtered value. 

\begin{figure}
\centering
\includegraphics[scale=0.3, trim=8.0cm 0.0cm 1.5cm 0.0cm]{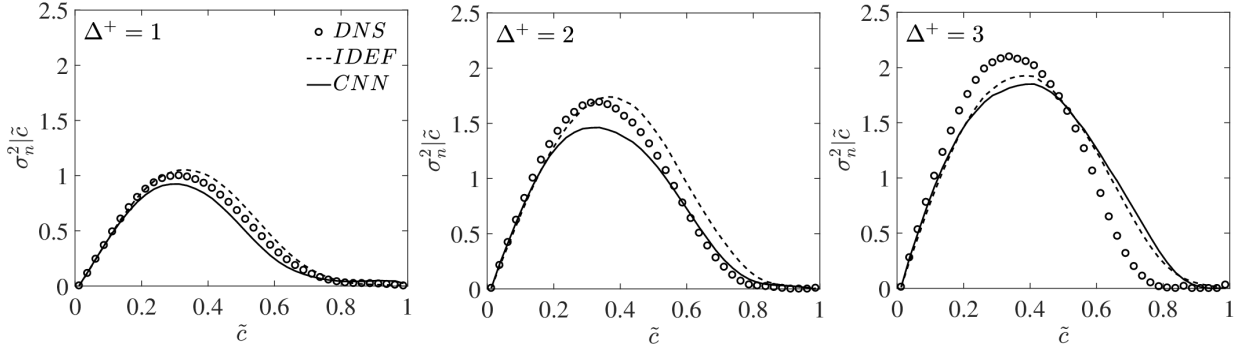}
\caption{Conditionally-averaged modelled progress variable variance against target results.}%sc01arh02
\label{fig:sigma2c}
\end{figure}

Figure \ref{fig:c2} (a)-(c) shows a comparison between the instantaneous filtered product $\widetilde{{c^*}^2}$ and the actual field  $\widetilde{c^2}$ as obtained on the LES mesh, for $\Delta ^+=1$, 2 and 3 respectively. The red line corresponds to $y=x$ and serves as a guideline. For all three filter widths, the CNN prediction of the filtered product is in good agreement with the target data. There is little scatter, and an almost linear correlation for all three cases. Figure \ref{fig:sigma2c} shows the results for the variance for $\Delta ^+=1$, 2 and 3 respectively. The conditional averages are normalised using the maximum variance value for $\Delta ^+=$1. The target results as obtained by explicitly filtering the data on the DNS mesh and sampling on the LES mesh are also shown. We additionally calculate the variance using a recently proposed method namely Iterative Deconvolution and Explicit Filtering (IDEF) \cite{nikolaou_prf_2018,nikolaou_ftc_2018_01} as a comparison. IDEF was shown in \cite{nikolaou_prf_2018,nikolaou_ftc_2018_01} to provide good estimates of the variance for all filter widths considered. In comparison to a popular classic gradient-based model, IDEF was shown in \cite{nikolaou_ftc_2018_01} to provide improved predictions hence it serves as a good benchmark. It is also important at this point to note that a good prediction of the filtered product $\widetilde{c^2}$ does not necessarily ensure a good prediction of the variance also. It is straightforward to show that for a given local percentage error $p_1$ in predicting $\widetilde{c^2}$, the corresponding error $p_2$ in predicting the variance is $p_2=p_1\widetilde{c^2}/(\widetilde{c^2}-\tilde{c} \tilde{c})$ i.e. the error in the variance is always larger since the fraction $\widetilde{c^2}/(\widetilde{c^2}-\tilde{c} \tilde{c})>$1.0 always. Therefore the aim is to keep $p_1$ as small as possible. This can be achieved by increasing the training data size and/or the complexity of the network. It is important to note however, that the optimum network structure which still minimises the optimisation error in deep networks is still and open area of research, and this depends strongly on the input data. In this study our aim was to keep the network structure to within reasonable complexity. 

For the smallest filter width, $\Delta ^+=1$, the progress variable variance prediction using CNNs as one may observe from the results in Fig. \ref{fig:sigma2c}, is in good agreement with the DNS data, with the performance almost matching that of IDEF. For $\Delta ^+=2$, and 3 which correspond to coarser meshes, the prediction is relatively poorer in the flame brush. This is primarily a result of the training data size since for larger filter widths this is smaller. As a result, fewer points in the flame brush are available for sufficient training. This results in a slight under-prediction of large $\widetilde{c^2}$ values as one may observe from the results in Fig. \ref{fig:c2} for these filter widths. However, the overall performance of the CNN-based deconvolution, a purely data-driven approach which does not involve any tunable, flow-regime-dependent parameters, is still remarkable for all three filter widths, and under the stringent testing scenarios in this study, almost matching that of IDEF across the entire range of $\tilde{c}$ values. It is expected that increasing the training data size, and/or the number of convolutional layers, the agreement can substantially improve, all of which are a subject of future research. 

\section{Combining with classic methods: filtered rate modelling}

In this section, we illustrate how CNN-based deconvolution can be used in conjunction with traditional modelling approaches for modelling additional terms such as the filtered progress variable reaction rate, which is a dominant term in the filtered progress variable transport equation. These include two popular approaches which require the progress variable variance as input, namely the Unstrained Flamelet approach (UF) and the Filtered Laminar Flame approach (FLF). In order to quantify the effect of the CNN-based modelled variance, the modelled rates are calculated using as input both the actual variance as obtained on the LES mesh, and also the modelled variance as obtained using the neural network. 

\vspace{0.5cm}

\noindent \textit{Unstrained flamelets:}  

\vspace{0.5cm}

In the classical UF modelling approach the filtered rate is calculated using, 

\begin{equation}\label{eq:ufmodel}
\overline{\dot{w}}_c(\underline{x},t)=\bar{\rho}(\underline{x},t)\int_{0}^{1} \frac{\dot{w}_{cL}(\zeta)}{\rho _L (\zeta)}\tilde{p}(\zeta;\tilde{c},{\sigma ^2})d\zeta
\end{equation}

\noindent where $\zeta$ is the sample space variable for $c$, $\dot{w}_{cL}$ is the laminar progress variable rate, and $\rho _{L}$ is the laminar density. The progress variable pdf is taken to be a $\beta$-function in accordance to usual practice in presumed pdf methods, 

\begin{equation*}
\tilde{p}(\zeta)=\left(\frac{1}{C}\right){\zeta}^{a-1}(1-\zeta)^{b-1}
\end{equation*}

\noindent where $C$ is a normalisation constant and where the parameters $a,b$ are chosen so that the filtered progress variable $\tilde{c}$ and the variance $\sigma ^2$ are recovered. These are given by: $a=\tilde{c} \left( 1/g-1 \right)$, and $b=(1-\tilde{c})\left( 1/g-1 \right)$, where $g={{\sigma}^2}/{\tilde{c}(1-\tilde{c})}$. 

Note that in practice, the integration as specified by Eq. \ref{eq:ufmodel} may be problematic since the progress variable pdf takes non-finite values in the case the pdf is bimodal. A way around this issue is to expand the integral in Eq. \ref{eq:ufmodel} and use the cumulative distribution function $\tilde{P}(\zeta;\tilde{c},{\sigma ^2})$ instead for calculating the mean of a variable $y$ as, 

\begin{equation}
\overline{y}(\underline{x},t)=
\frac{\bar{\rho}(\underline{x},t)}{\rho _L(1)}y_L(1) \\
-\bar{\rho}(\underline{x},t) \int_0^1 \frac{1}{\rho _L(\zeta)}\left( \rho _L(\zeta) \frac{d y_L(\zeta)}{d \zeta}-y_L(\zeta) \frac{d \rho _L (\zeta)}{d \zeta} \right ) \tilde{P}(\zeta;\tilde{c},{\sigma ^2}) d \zeta
\end{equation}

\noindent where the derivatives of $y$ with respect to $\zeta$ are typically well-defined for quantities of interest, and so is the cumulative distribution function $\tilde{P}$. 

\vspace{0.5cm}

\noindent \textit{Filtered laminar flame approach:}  

\vspace{0.5cm}

In the FLF approach \cite{moureau_cnf_2010,nambully_cnf_2014}, the filtered rate is calculated from look-up tables constructed using filtered laminar profiles of a canonical flame solution e.g. a 1D unstrained flame. The filtered rate is calculated using, 

\begin{equation}
\bar{\dot{w}}_c(\underline{x},t)=\bar{\rho}(\underline{x},t)\int_{0}^{1}\frac{\dot{w}_{cL}(\zeta)}{\rho _L(\zeta)}\tilde{p}_L(\zeta;\tilde{c};\sigma ^2)d\zeta
\end{equation}

\noindent This process essentially corresponds to a filtered laminar flame pdf approach \cite{moureau_cnf_2010,nambully_cnf_2014}. The filtered laminar flame pdf is given by,

\begin{equation*}
\tilde{p}_L(\zeta;\tilde{c};\sigma ^2)=\frac{\rho _L(\zeta)}{\bar{\rho} ^{\Delta _L} _L} \left| \frac{d \zeta}{dx} \right| ^{-1} G\left({x} ^{\Delta _L} _L-x(\zeta) \right)
\end{equation*}

\noindent In the FLF approach, the laminar flame filter size $\Delta _L$ which is used to parameterise the laminar flame pdf, is generally smaller than the actual LES filter size $\Delta$. The laminar flame filter $\Delta _L$, and the laminar flame position $x_L$, are chosen so that the progress variable and its variance as obtained from the LES, match the corresponding 1D-filtered laminar flame values i.e. $\tilde{c}(\underline{x},t;\Delta)=\tilde{c}_L(x_L;\Delta _L)$, and $\sigma ^2(\underline{x},t;\Delta)=\sigma ^2 _L(x_L;\Delta _L)$. This ensures that the filtered-flame pdf recovers the LES mean and variance values. In practice, an unstrained laminar flame is filtered for a wide range of filter widths in order to obtain a table of values of $\tilde{c} _L$, and $\sigma ^2 _L$. The LES values $\tilde{c}$ and $\sigma ^2$ are then used in order to determine the corresponding filter width $\Delta _L$ and spatial position $x_L$ in the table which match the filtered laminar flame values. The mean rate is then calculated by filtering the laminar flame profile at $\Delta _L$ and obtaining its value at $x_L$ \cite{moureau_cnf_2010,nambully_cnf_2014}. 

\begin{figure}
\centering
\includegraphics[scale=0.32, trim=11.0cm 0.0cm 1.5cm 0.0cm]{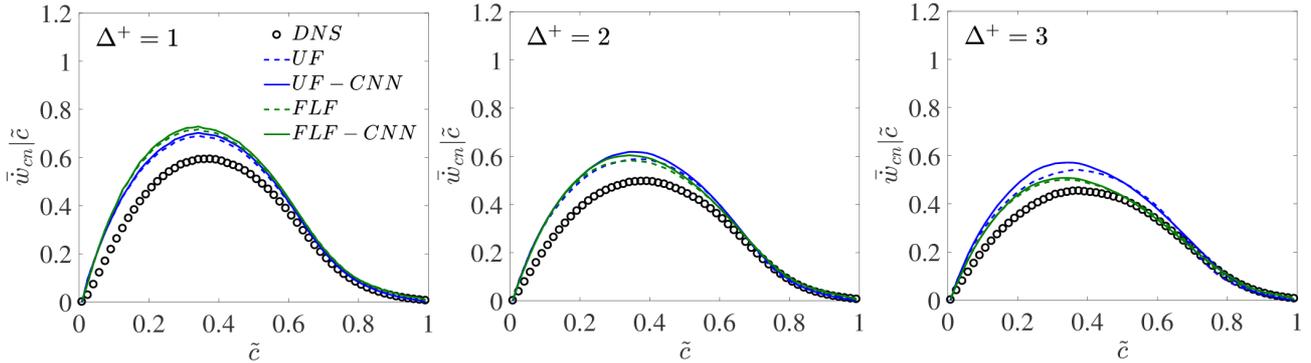}
\caption{Conditional modelled progress variable reaction rate.}%sc01arh02
\label{fig:cond_wc_all}
\end{figure}

Figure \ref{fig:cond_wc_all} shows the conditional modelled rates using the above two classic flamelet methods for case C. For each filter width, the modelled rates are obtained by using the actual variance to parameterise the progress variable pdfs as obtained on the LES mesh. These are denoted as UF and FLF for the unstrained flamelet and filtered laminar flame approaches respectively. The CNN-modelled variance is also used to parameterise the corresponding progress variable pdfs of the two flamelet models, and these predictions are denoted by UF-CNN, and FLF-CNN. The target filtered rate as obtained by explicitly filtering on the DNS mesh and then sampling onto the LES mesh is also shown. A fine point to note is that the target results only serve as a guideline-our aim here is to examine any discrepancies between the modelled rates as a result of any differences in the modelled and actual variance. For all three filter widths, the conditional modelled rates are almost identical for both flamelet approaches across the entire range of $\tilde{c}$ values. The modelled rates are also found to be in good agreement with the target DNS results, though there is a slight over-estimation of the filtered rate in the region $0.1<\tilde{c}<0.6$ approximately-this is owed to the high turbulence level where the laminar flamelet concept may no longer hold for this case. Indeed, in \cite{nikolaou_cst_2015} it was shown that case C exhibits a highly patchy flame surface. Nevertheless, the main point of this section is that the CNN-based modelled variance does not introduce any significant bias in the two modelling approaches. Hence such neural-network deconvolution-based modelling methods can be safely used in conjunction with classic modelling approaches for modelling additional terms in the governing equations.    

\section*{Conclusions}

A data-driven method using deep convolutional neural networks is proposed for modelling purposes in large eddy simulations of reacting flows. A convolutional network has been trained to perform deconvolution on the filtered density and the filtered density-progress variable product using data from direct numerical simulations of turbulent freely-propagating turbulent premixed flames. 

The network is able to recover good approximations of the un-filtered fields which are then used to model the progress variable variance. The network requires no explicit knowledge of filtering kernel, and is shown to be relatively insensitive to filter-width variations, in contrast to the majority of classic models. In principle, the method described in this study is general and can be applied to databases generated from larger-scale direct simulations to increase the training data size, hence developing networks of increased accuracy, for any variable of interest.

Data-driven methods in general, have the potential to re-direct and revolutionise on-going modelling efforts both in computational fluid dynamics but also in other systems of non-linear conservation laws. In contrast to traditional modelling approaches, an explicit knowledge of the underlying physics is not required, however it is important to note that the performance of such methods is limited by the amount and quality of the training data. 

%--------------------------------------------------------------------

\vspace{1cm}

\textbf{Competing interests:} The authors declare they have no competing interests.

%-------------------------------------------------
\newpage
\section*{Appendix}

\subsection*{Network structure}

A CNN usually consists of convolutional and sub-sampling layers accompanied by fully connected layers. Each of the convolutional layers of the CNN can have $K$ number of filters (kernels). An essential aspect of the CNN is the size of the filters, which can identify the locally connected structure and in-turn convoluted with the input to create $K$ feature maps. Each of the generated $K$ feature maps can then be sub-sampled using min or max pooling for a defined region, typically between 2-5 points. Furthermore, another important part of the CNN is the addition of a bias parameter and the application of a linear or non-linear activation function for each feature map. The use of bias and the activation function can be applied either before or after the sampling of the feature maps. The training of the CNN can be performed using the back-propagation algorithm \cite{lecun1990handwritten}. 

\begin{figure}[h!]
\begin{center}
  \includegraphics[scale=0.33]{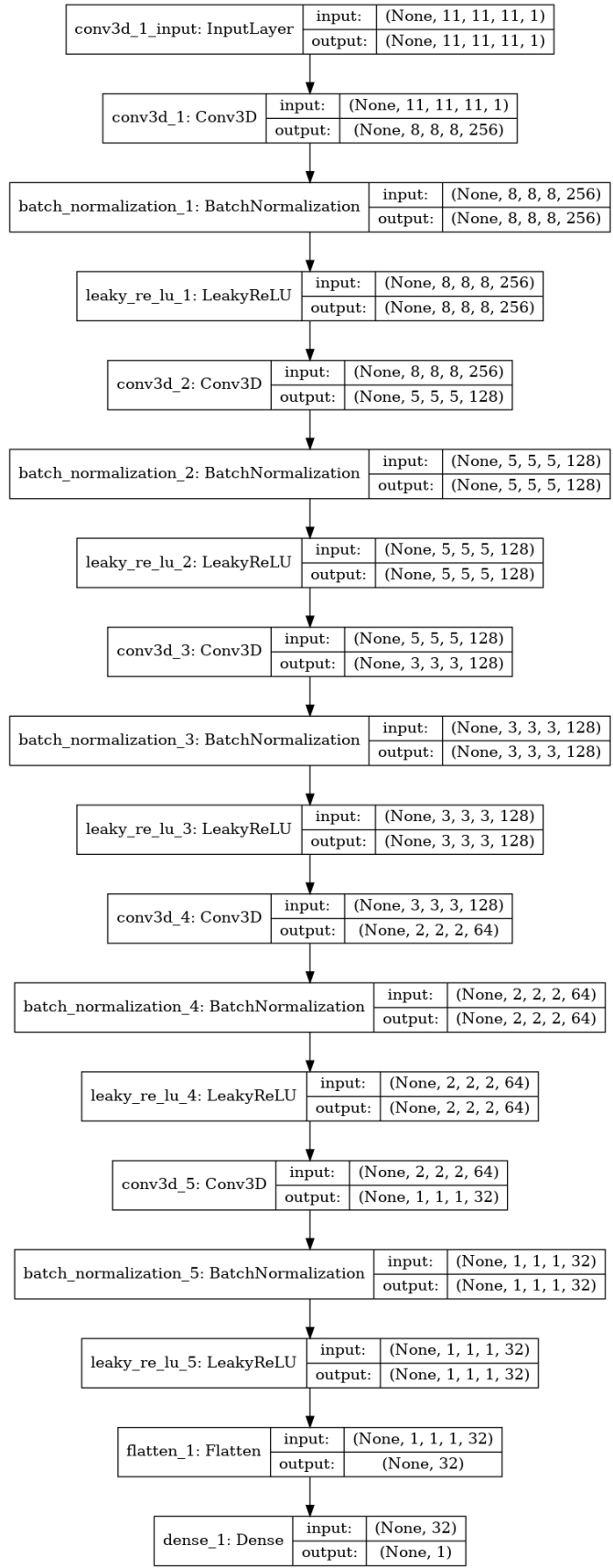}
  \caption{The structure of the convolutional network.}
  \label{fig:cnn_struct}
  \end{center}
\end{figure}

Figure \ref{fig:cnn_struct} shows the structure of the network used for the deconvolution. The input layer consists of a set of $11^3$ points holding the filtered values in the halo cube. The output from the first convolutional layer is an $8^3$ set of features for each of the 256 kernels used. Following this, a leaky Rectified Linear Unit function (RELU) is applied. In the second convolutional layer, the output from the RELU is convoluted using 128 kernels, resulting in an output set of $5^3$ features for each of the 128 kernels. The process is repeated with convolutional, RELU and normalisation layers. During the training phase, the weights of all kernels in each convolutional layer are adjusted so as to minimise the mean-squared error between the deconvoluted and actual field. In the end, a total of 32 features are extracted which are connected to a single node having a linear activation function and resulting in a single output namely the deconvoluted field. 

The total size of the training data for each case depends on the size of the LES mesh. In particular, for an LES mesh having $N_x, N_y, N_z$ points in space and $N_t$ datasets in time, the total size of the training data is $N_x \cdot N_y \cdot N_z \cdot N_t \cdot N_h^3$ where $N_h$ is the size of the halo cube around each point on the LES mesh. Therefore depending on the size of the DNS database the training data size can become significant. The open-source Python-based library ``Tensor Flow" was used for developing the network \cite{tensor_flow}, which was designed to run on Graphics Processing Units (GPUs), thus enabling accelerated training times for such large data sets.  

\clearpage
\newpage

\end{document}